# Bound polaron in CuCl and GaAs Quantum Dots

H. Satori [1, 2], M. Harti [1, 2], N. Chenfour [2], F. Assaoui [3]

*(1) : UFR Informatique et Nouvelles Technologies d'Information et de Communication B.P. 1796, Dhar Mehraz Fès Morocco.*
*(2) : Département de Mathématiques et Informatique, Faculté des Sciences, B.P. 1796, Dhar Mehraz Fès, Morocco.*
*(3) : High Energy Physics Laboratory, Faculty of Sciences, Md V University Agdal -Rabat, Morocco.*

**ABSTRACT:**
 We calculated the binding energy of a polaron bound to a hydrogenic donor impurity located in a spherical quantum dot by means of a variational and numerical technique for finite potential models. The polaronic effect has been considered taking into account the ion-phonon coupling under the Lee Low Pines approach. The results show that the binding energies are drastically affected by the dot radius, the potential barrier height and the polaronic effects.

## I. INTRODUCTION

It has been increasing interest for quantum dots (Q.D) structures from the physical point of view and for potential electronic and optical device application [1-4]. The presence of the impurity state in the Q.D system plays a fundamental role in some physical properties such as the electronic mobility, electronic transport and optical properties. It is found that the strong electronic confinement in these structures modifies the ground state energy and the impurity binding energy.

In polar semiconductors [5] such as GaAs, the interaction of the electron with the optical phonons is very important to understand the optical absorption spectra. The polaronic effect is therefore an interesting subject in the low dimensional system. Recently, many investigations have been advanced for electronic properties in a polar Q.D semiconductor taking into account the electron-phonon interaction [6-10].

Many theoretical and experimental studies have dealt with polaronic effect on the binding energy in Q.Ds without taking into account the interaction between the impurity ion and the phonons. Indeed, for semiconductor materials with intermediate and strong coupling, the ion-phonon interaction can not be neglected [10-12].

The aim of the present paper is to study the effect of interaction of charge carriers (electron and ion) with the longitudinal optical (LO) phonons as well as the surface optical (SO) phonons on the binding energy of a donor impurity in a spherical Q.D embedded in a dielectric matrix. The charge carriers-phonon interactions are described by using a variational approach and Lee-Low-Pines transformation [9], which is suitable for the weak and intermediate coupling. In next sections, we present the basic theory of our calculations, and we present and discus our results.

## II. BASIC THEORY

Within the effective mass approximation, the Hamiltonian of a hydrogenic impurity system confined in a polar spherical Q.D of radius R, embedded in a dielectric matrix of constant $\varepsilon_d$, interacting with different optical modes (LO and SO) can be written as:

$$H = H_e + H_{ph} + H_{e\text{-}ph} + H_{ion\text{-}ph} \tag{1}$$

The electronic part $H_e$ is given by:

$$H_e = \frac{P^2}{2m^*} - \frac{e^2}{\varepsilon_\infty |r - r_0|} + V_{conf}(r) \quad , \tag{2}$$

where P, **r** and m* are momentum, position and effective mass of the electron, respectively. $\varepsilon_\infty$ stand for the high frequency optical dielectric constant inside the Q.D material, $\mathbf{r}_0$ is the impurity position measured from the center of the spherical Q.D. The confining potential $V_{conf}(r)$ for a dot is defined as:

$$V_{conf}(r) = \begin{cases} 0, & r \leq R \\ V_0, & r > R \end{cases}$$

$V_0$ being the height of the potential barrier, which is the difference of the conduction band offsets of the dot material and the surrounding material. The phonon field Hamiltonian $H_{ph}$ is written as:

$$H_{ph} = \sum_{\ell m k} \hbar\omega_{LO} b^+_{\ell m}(k) b_{\ell m}(k) + \sum_{\ell m} \hbar\omega_{SO} a^+_{\ell m} a_{\ell m} \qquad (3)$$

The first and the second term in equation (3) are the free phonons Hamiltonian related, respectively, to LO and SO phonons. $b^+_{\ell m}(k)$ and $a^+_{\ell m}$ are the creation operators for the LO and SO phonons, respectively. $b_{\ell m}(k)$ and $a_{\ell m}$ are the annihilation operators for the LO and SO phonons, respectively and k and $\hbar\omega_{LO}$ represent, respectively, the wave number and the energy for the LO phonons mode.

The third term and the last term in equation (1) represent, respectively, the electron-phonon interaction Hamiltonian ($H_{e-ph}$) and the ion-phonon interaction Hamiltonian ($H_{ion-ph}$) and they can be written as follows: [9].

$$H_{e-ph} = \sum_{\ell m k} f_\ell^{LO}(k)\left[S^{LO}(r,\theta,\varphi) b_{\ell m}(k) + h.c\right] + \sum_{\ell m} f_\ell^{SO}\left[S^{SO}(r,\theta,\varphi) a_{\ell m} + h.c\right] \qquad (4)$$

and

$$H_{ion-ph} = -\sum_{\ell m k} f_\ell^{LO}(k)\left[S^{LO}(r_0,\theta_0,\varphi_0) b_{\ell m}(k) + h.c\right] - \sum_{\ell m} f_\ell^{SO}\left[S^{SO}(r_0,\theta_0,\varphi_0) a_{\ell m} + h.c\right] \qquad (5)$$

Where

$$S^{SO}(r,\theta,\varphi) = \begin{cases} \left(\frac{r}{R}\right)^\ell Y_\ell^m(\hat{r}) & \text{for } r \leq R \\ \left(\frac{R}{r}\right)^{\ell+1} Y_\ell^m(\hat{r}) & \text{for } r > R \end{cases}, \quad S^{LO}(r,\theta,\varphi) = \begin{cases} j_\ell(kr) Y_\ell^m(\hat{r}) & \text{for } r \leq R \\ 0 & \text{for } r > R \end{cases}.$$

The functions $j_\ell(x)$ and $Y_\ell^m(\hat{r})$ [12] are the spherical Bessel functions of the $\ell$'th order and the spherical harmonics, respectively. $f_\ell^{LO}(k)$, $f_\ell^{SO}$ are the coupling coefficients for Lo and SO modes. $\varepsilon_0$ is the low frequency dielectric constant inside the spherical dot.

Following Lee-Low Pines theory, the trial function is written as the product form of the electronic state $|\psi_e\rangle$, the zero phonon state $|0\rangle$ and the unitary operator U [9, 13]:

$$|\psi\rangle = U|\psi_e\rangle|0\rangle. \qquad (6)$$

The expectation value of the energy in the ground state is given by:

$$E = E_e + E_{ph}^{LO} + E_{ph}^{SO} \qquad (7)$$

where

$$E_e = \frac{\langle\psi_e(r)|-\nabla^2 - \frac{2\varepsilon_0}{\varepsilon_\infty|r-r_0|} + V_{conf}(r)|\psi_e(r)\rangle}{\langle\psi_e(r)\|\psi_e(r)\rangle}, \quad E_{ph}^{LO}(\alpha) = -\sum_{\ell m k}\frac{|f_\ell^{LO}(k)|^2 |\rho^{LO}|^2}{\hbar\omega_{LO}} \text{ and } E_{ph}^{SO}(\alpha) = -\sum_{\ell m}\frac{|f_\ell^{SO}|^2 |\rho^{SO}|^2}{\hbar\omega_{SO}} \qquad (8)$$

with

$$\rho^{LO} = \frac{\langle\psi_e(r)|S^{LO}(r)-S^{LO}(r_0)|\psi_e(r)\rangle}{\langle\psi_e(r)\|\psi_e(r)\rangle} \text{ and } \rho^{SO} = \frac{\langle\psi_e(r)|S^{SO}(r)-S^{SO}(r_0)|\psi_e(r)\rangle}{\langle\psi_e(r)\|\psi_e(r)\rangle}.$$

$E_{ph}^{LO}(\alpha)$ and $E_{ph}^{SO}(\alpha)$ are the contributions of the two phonon modes (LO and SO) to the ground state energy of the bound polaron in spherical Q.D. Their expressions and its of the electronic part $E_e$ (in the general case of $r_0 \neq 0$) are complicated and can be evaluated only numerically.

We defined the energy system without the Coulomb interaction as:

$$E_{sub} = E_e^0 + E_{ph}^{LO}(\alpha = 0) + E_{ph}^{SO}(\alpha = 0) \qquad (25) \quad (24)$$

where

$$E_e^0 = \frac{\langle \psi_e(r, \alpha=0) | -\nabla^2 + V_{conf}(r) | \psi_e(r, \alpha=0) \rangle}{\langle \psi_e(r, \alpha=0) \| \psi_e(r, \alpha=0) \rangle}. \qquad (9)$$

The variational electronic wave function was taken as a product form $|\psi_e\rangle = \phi(r) \exp(-\alpha|\mathbf{r}-\mathbf{r_0}|)$; $\alpha$ is a variational parameter obtained by minimising the total energy of the system.

Where

$$\phi(r) = \begin{cases} \dfrac{\sin(\xi r)}{r} & r \leq R \\ \dfrac{\sin(\xi R)}{r} \exp \chi(R-r) & r > R \end{cases}, \qquad (10)$$

$$\xi = (2m_1^* E_0)^{1/2}/\hbar \quad \text{and} \quad \chi = [2m_2^*(V_0 - E_0)]^{1/2}/\hbar \qquad (11)$$

$m_1$ and $m_2$ are the effective masses inside and outside the dot, respectively. The ground level $E_0$ is determined by using the appropriate current conserving boundary condition for the wave function at the interface. It must satisfy the following relation.[15]

$$-\xi = \left\{ \frac{m_1}{m_2}\chi - \left(1 - \frac{m_1}{m_2}\right)\Big/R \right\} \tan(\xi R) \cdot \qquad (12)$$

the smallest radius for the existence of a bound state can be obtained from Eq. (12),

$$R_S = \left\{ \frac{\pi^2\hbar^2}{8m_1 V_0} + \frac{\hbar^2}{2m_2 V_0}\left(\frac{m_2}{m_1} - 1\right)^2 \right\}^{1/2}.$$

The binding energy $E_b$ of the system is defined as the energy difference between the bottom of the electronic conduction band without the Coulomb interaction $E_{sub}$ and the ground state energy of the polaron E, taking into account the polaronic effect in both situations. The polaron energy E is given by the minimisation of the expectation value of the energy with respect to the variational parameters $\alpha$, i.e.

$$E_b = E_{Sub} - \min_{\alpha}\{E\} \qquad (13)$$

We note that the correction given by the SO phonon vanished for an impurity located at the center of the dot ($r_0 = 0$), this result is in accordance with that obtained for donor like exciton [13]. This correction does not vanish for $r_0 \neq 0$ and yield only very small effects in comparison with confined LO phonons, which have a dominant role.

## III. RESULTS AND DISCUSSIONS

Recent experimental techniques have made possible the fabrication of a quantum dot with size of a few nanometers and a variety of shapes such as spherical and rectangular shapes. In our study, we have chosen for the numerical results the weakly polar material GaAs ($\alpha_{GaAs} = 0.06$) and the CuCl material whose electron-phonon coupling is intermediate ($\alpha_{CuCl} = 2.45$). The values of physical parameters used for the calculation of the binding energy of a bound polaron in spherical Q.Ds are: $\varepsilon_0 = 7.9$, $\varepsilon_\infty = 3.61$, and $m^*/m_0 = 0.5$ for CuCl, where $m_0$ is the free electron mass. $\varepsilon_0 = 12.5$, $\varepsilon_\infty = 10.9$, and $m^*/m_0 = 0.06$ for GaAs. In all what follows, the results are displayed in atomic units (a.u) of length $a^* = \hbar^2\varepsilon_0/m^*e^2$ and energy $R^* = m^*e^4/2\hbar^2\varepsilon_0^2$, i.e., respectively $8.36\,\text{A}°$, $109.0\,\text{meV}$ for CuCl, $100.2\,\text{A}°$ and $5.75\,\text{meV}$ for GaAs. The nonpolar medium, which surrounds the dot, is characterised by the dielectric constant $\varepsilon_d = 2.25$ and the height potential barrier is taken to be $V_0 = 20\,R^*$ for both materials.

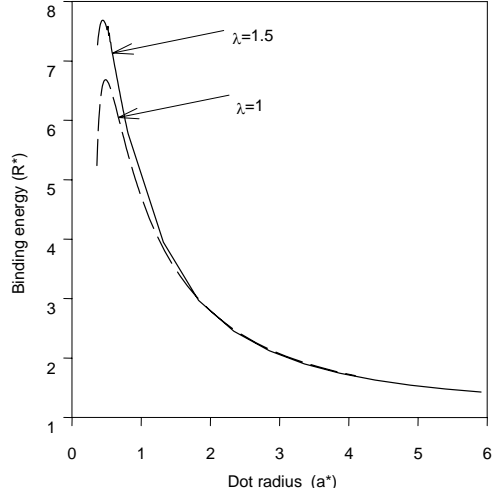

**FIG.1:** Binding energy of a donor impurity in a spherical GaAs Q.D with phonons correction as a function of the dot radius, taking the ratio $\lambda = m_2/m_1$ as a parameter; i.e. $\lambda=1$ and $\lambda=1.5$.

In the case of GaAs Q.D, we have shown in figure-1 the binding energy of an impurity placed at the center versus the dot radius $R(a^*)$ for two values of the ratio $\lambda = m_2/m_1$ ($\lambda=1$ and 1.5) and for a fixed value of the barrier ($V_0 = 20 R^*$) taking into account the charge carriers phonon interactions. As we can note from this figure, for dot radius such as $R \geq a^*$, the effect of the ratio $E_u$ on the binding energy is negligible and the binding energies of the two values of $\lambda$ ($\lambda=1$ and $\lambda=1.5$) have the same numerical values, since the major party of the electronic density is localised into the dot. On the contrary, for dot radius $R<a^*$, the binding energy presents a maximum which increases with the ratio $\lambda$. It is interesting to mention that for large dot radius $R>a^*$, the effective mass approximation theory agrees with the other models of the first principle [13]. Since, in the region of the strong confinement, the results obtained within the effective mass approximation overestimate those of the first Principe calculation. Hence, the penetration of wave function in the surrounding material is more significant and depends strongly on the height of the potential barrier and the anisotropy of the effective mass.

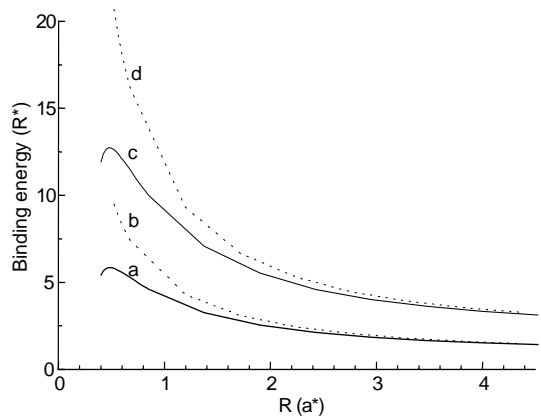

**FIG.2-1**: Binding energy of a donor impurity in a spherical CuCl Q.D as function of the dot radius for infinite potential barrier (dotted curves), and for the finite potential model (full curves). The curves (a, b) and (c, d) are, respectively, the binding energies without phonons, with phonon.

In Fig-2-1 we present the binding energy of a hydrogenic impurity placed at the center of spherical CuCl Q.D embedded in a dielectric matrix versus the dot radius for both the infinite potential model (dotted curves) and the finite potential barrier model (full curves) with a fixed value of $\lambda=1$. Curves (a) and (b) correspond to the case without phonons and (c) and (d) represent the case with phonons. For the finite barrier model (curves (a) and (c)) as the dot radius decreases, the binding energy increases, reaches a maximum value and then decreases to its barrier material value which occurs at the radius threshold $R_s$.

For Q.D of large radius, the binding energy converges to the bulk value of material CuCl. Whereas for the infinite well model, as the dot size decreases, the binding energy increases monotonically from its bulk value. For Q.D of radius R larger than $R_0 \cong 2 \times a*$, the effect of the confinement potential ($V_0 \cong 20 \times R*$) is negligible.

For large value of the size of the dot, the penetration of the wave function in the surrounding material is weak because of a small electronic confinement. As a consequence, the numerical values of the binding energy for both the infinite and finite models are nearly similar.

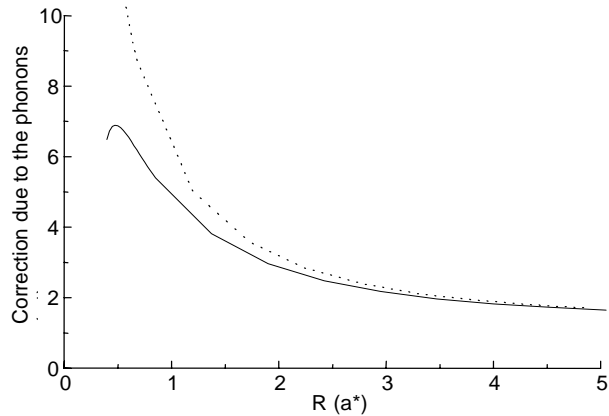

**FIG.2-2**: Correction due to the phonons effect for the CuCl Q.D as function of the dot radius for both infinite potential model (dotted curve) and finite potential model (full curve).

While for $R < R_0$, this effect is more pronounced as the dot becomes thin. The comparison of curves (a) and (c) in the realistic case reveals that the correction induced by the
charge carriers-phonons interaction increases as the electronic confinement increases. As we can note this effect is more significant for the case of infinite barrier model. These results are clearly observed in figure 2-2 where we have shown the correction in binding energy due
to the polaronic effect in the case of CuCl Q.D. The large magnitude of correction given by the phonons in a small dot is due to an existence of an electron even if the dot radius becomes very small, which yields the stronger coupling between the charge carriers and phonons.

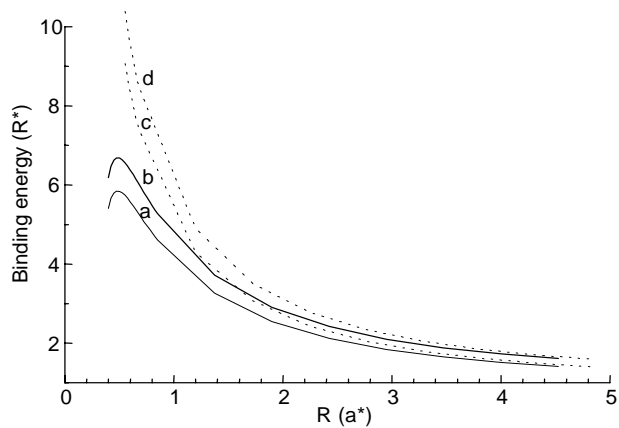

**FIG.3-1:** Binding energy of a donor impurity in a spherical GaAs Q.D as a function of the dot radius. The curves a, b, c and d represent the same situation as in Fig-2-1.

In Figure 3-1 and Figure 3-2, we have reproduced the case for the weakly polar material Q.D GaAs. The same behaviour is exhibited as in figures 2-1 and 2-2. The comparison between Figure 2-2 and Figure 3-2 shows that the correction due to the phonons is more pronounced for the more polar Q.D ( CuCl ) than for the weakly polar Q.D (GaAs). This originates from the increasing localisation of the wave function with strong electron-(ion-) phonon coupling.

In order to study the nature of polaronic effects on a polaron in GaAs and CuCl Q.Ds, we have used the L.L.P variational method suitable for the weak and intermediate coupling. Then, it is important to note that in the case of CuCl Q.D, the large magnitude of the correction given by the phonon modes for a smaller dot, which yields to the enhancement of coupling between electron and phonon [11], limited our calculation to a certain values of the dot radius, relatively large. Thus, we expect that for dot radius grater than 1 a*, where the effective mass approximation is valid, our method can give reasonable results because the electorn-phonon coupling remains moderate in this region.

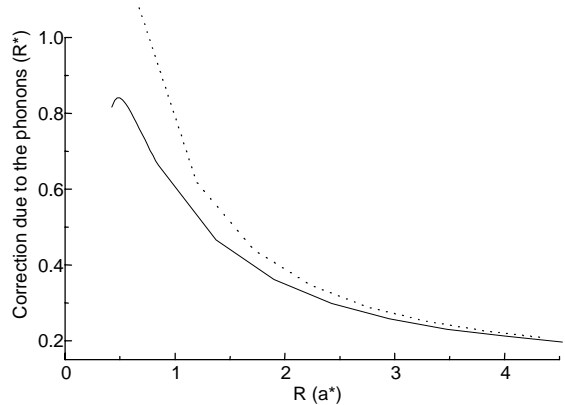

**FIG.3-2**: Correction due to the phonons effect for the GaAs Q.D as function of the dot radius for both infinite potential model (dotted curve) and finite potential model (full curve).

## IV. CONCLUSION

We have calculated the binding energy of a polaron bound to a hydrogenic impurity located at the center of a spherical Q.D in both weakly and more polar materials in the infinite and finite potential model cases. The interaction between the electron-phonons and the ion-phonons coupling has been taken into account in our study using a modified Lee-Low-Pines variational treatment. We have found that the polaronic effect influences drastically the binding energy and increases with the increasing electronic confinement and the electron-phonon coupling strength.

**Acknowledgements** : This work has been done with the financial support by ″Sat.FEM/06″ .